\pdfoutput=1
\documentclass[12pt]{iopart}

\usepackage{graphicx}
\usepackage{color}
\begin{document}

\title[Brownian fluctuations of kinetic energy under Lorentz force.]{Brownian fluctuations of kinetic energy under Lorentz force.}

\author{Pedro V. Paraguass\'{u}}

\address{Departamento de F\'{i}sica, Pontif\'{i}cia Universidade Cat\'{o}lica\\ 22452-970, Rio de Janeiro, Brazil}

\ead{paraguassu@aluno.puc-rio.br}
\vspace{10pt}

\begin{indented}
\item[]June 2023
\end{indented}

\begin{abstract}
    In stochastic thermodynamics, significant attention has been given to studying the statistical behavior of thermodynamic quantities such as heat and work. However, fluctuations in other quantities, such as kinetic energy and internal energy, can also exhibit intriguing statistical properties. In this study, we investigate the fluctuations of kinetic energy within an initially equilibrated underdamped Brownian particle subsequently exposed to a Lorentz force, comprising both electric and magnetic fields. Providing insights through the examination of the characteristic function, central moments, and kinetic energy distribution.
 
\end{abstract}

%
%
%
%
%

\section{Introduction}

Kinetic energy is the energy that a system possesses due to its motion, whether it is translational or rotational \cite{nussenzveig2013curso}. In stochastic systems, the motion occurs randomly due to noise. As a consequence, the kinetic energy for stochastic systems will also be random, being a functional of velocities. Understanding these Brownian fluctuations in kinetic energy helps to comprehend the fluctuations of heat within Stochastic Thermodynamics, since the kinetic energy always contributes to the heat \cite{paraguassu2022effects}.

The framework of Stochastic Thermodynamics \cite{peliti2021stochastic, oliveira2020classical, sekimoto2010stochastic} has emerged as a valuable approach to comprehend energy transfer within systems. It investigates fluctuating quantities such as heat, work, and entropy, which have been extensively studied in the literature \cite{paraguassu2022effects,paraguassu2023heat,paraguassu2023heat2,salazar2020work,chatterjee2011single,chatterjee2010exact,imparato2008probability,saha2014work,joubaud2007fluctuation, pires2023optimal}. These fluctuations have a wide range of applications \cite{peliti2021stochastic}, including stochastic thermal machines \cite{holubec2021fluctuations,martinez2016brownian,blickle2012realization,paneru2020efficiency}, where the interplay between efficiency and power is influenced by fluctuations \cite{holubec2021fluctuations,fernando2022thermodynamics,verley2014unlikely,schmiedl2007efficiency}. The scope of Stochastic Thermodynamics has expanded significantly, encompassing various domains from quantum systems \cite{campisi2016quantum, esposito2009nonequilibrium,funo2018path, oliveira2020classical} to relativistic systems \cite{koide2011thermodynamic,pal2020stochastic,paraguassu2021heat}. This growing range of applications demonstrates the increasing relevance and versatility of Stochastic Thermodynamics in understanding and analyzing diverse physical systems. It is with this framework that we aim to analyze the kinetic energy of a Brownian system.

But why calculate the kinetic energy? It is a term present in the formula for the heat, even in the overdamped case \cite{paraguassu2022effects}. Additionally, it is a component of the total energy of the Brownian particle. Furthermore, for instance, in the case of a free particle, the heat is equal to the kinetic energy, regardless of whether the system is overdamped or underdamped \cite{paraguassu2023heat,paraguassu2022effects}. Therefore, understanding the Brownian fluctuations of kinetic energy helps us comprehend the energetics of such systems. While the dynamics of velocity have been extensively investigated \cite{czopnik2001brownian,lemons1999brownian,hou2009brownian}, the fluctuations of the kinetic energy has received relatively little attention. However, it can yield non-trivial probability distributions that warrant further exploration.

Here, we are interested in investigating the distribution and fluctuations of kinetic energy for a two-dimensional Brownian particle in the underdamped regime. Initially, the particle is in free equilibrium, and it is later influenced by a static magnetic and electric fields resulting from the Lorentz force. The model of a magnetic field along with an electric field acting on a charged Brownian particle is a classic model widely used to describe the diffusion of electrons and heavy ions in plasma \cite{lemons1999brownian,hou2009brownian,czopnik2001brownian,kurcsunoglu1962brownian,lucero2020brownian,butanas2022brownian,kurcsunoglu1963brownian,liboff1966brownian}. Our study introduces a new perspective by emphasizing the energetics, as opposed to solely examining the dynamics, thus offering a novel insight into these systems.

Moreover, the impact of magnetic fields on stochastic systems has been previously explored in the literature, particularly in the context of the Bohr-Van Leeuwen theorem \cite{lucero2020brownian,matevosyan_lasting_2023,pradhan_nonexistence_2010,vidal-urquiza_dynamics_2017}. Furthermore, thermodynamic functionals have been investigated in the works \cite{chatterjee2011single} and \cite{saha2008nonequilibrium}, where calculations were performed for work and internal energy distributions in the presence of a magnetic field. Our analysis builds upon these studies.

To obtain the kinetic energy distribution, we employ path integrals for our stochastic systems \cite{wio2013path,chaichian2018path,suassuna2021path,moreno2019conditional}. The results encompass the characteristic function, associated central moments, and the distribution of kinetic energy. These results are presented analytically and compared with numerical simulations, demonstrating agreement between the two approaches. We find that with only the magnetic field, the kinetic energy is not affected, a consequence of the Bohr-Van Leeuwen theorem, while for the electric field case the kinetic energy is affected and driven by the electric field. Besides, by considering the electric field together with the magnetic field, the magnetic field becomes a permanent effect, persisting in the motion of the particle, and thus affecting the fluctuations of the kinetic energy.

The remaining of this manuscript is organized as follows: In Sec.~\ref{model}, we introduce our model and specify its thermodynamics properties in Sec.~\ref{probab} we describe the probabilistic features of the system, its conditional probability distribution and its asymptotic distribution, for the three cases considered, the pure magnetic field, the pure electric field, and the combination of both. In Sec.~\ref{Cfun} we calculate the characteristic function and the central moments of the kinetic energy, dicussing these findings and comparing with the different cases, and in Sec.~\ref{Dist} we find the kinetic energy distribution and discuss our results. Conclusions and final remarks are addressed to Sec.~\ref{conclusion}.

\section{Model and Thermodynamics}
\label{model}

Our starting point is a two-dimensional underdamped Langevin equation with a constant magnetic and an electric force,
\begin{eqnarray}
    m \dot v_x(t) = - \gamma v_x(t) + \eta_x(t)-\Gamma v_y(t)+E_x,\nonumber\\ m \dot v_y(t) = - \gamma v_y(t) + \eta_y(t) + \Gamma v_x(t) +E_y,
    \label{langevin}
\end{eqnarray}
which describes the motion of a two-dimensional Brownian particle diffusing in a fluid under the influence of the Lorentz force $\vec{F}=(-\Gamma v_y+E_x,\Gamma v_x+E_y)$ \cite{nussenzveig2013curso}. This force can be interpreted as the magnetic force resulting from a static magnetic field perpendicular to the plane of motion, together with an electric force coming from a linear electrostatic potential. In this specific case, we have $\Gamma = qB_0$, where $q$ represents the charge, and $B_0$ represents the magnitude of the static magnetic field, and $E_i=q\epsilon_i$, where $\epsilon_i$ is the constant electric field. The thermal noises $\eta_x(t)$ and $\eta_y(t)$ are defined by $\langle\eta_{{x,y}}(t) \rangle=0$, and $\langle \eta_i(t)\eta_j(t')\rangle = 2\delta_{i,j}\gamma T\delta(t-t')$, with $i,j=x,y$, where $T$ is the temperature of the fluid.
This system is mathematically similar to the overdamped system investigated in \cite{chatterjee2011single}. However, here, we are in the underdamped regime, with the addition of an electric field along with the magnetic field.

Let's analyze the thermodynamics of the system, starting with the concept of heat.
The heat exchanged between the time interval $t\in[0,\tau]$ will be given by \cite{peliti2021stochastic,paraguassu2022effects}
\begin{equation}
   Q= \int_{0}^{\tau} \left(-\gamma v_x + \eta_x\right) v_x dt + \int_{0}^{\tau} \left(-\gamma v_y + \eta_y\right) v_y dt 
\end{equation}
where the Stratonovich prescription is assumed in the products of stochastic variables, as it is the correct one for Stochastic Thermodynamics \cite{bo2019functionals}. By using the Langevin equation given by Eq.~\eref{langevin} in the definition of heat, we obtain
\begin{equation}
    Q= \Delta K  - \int_0^\tau \left(E_x v_x +E_y v_y\right)dt. 
\end{equation}
It is important to note that the heat now depends on the velocities degrees of freedom, associated with $x(t)$ and $y(t)$. This introduces a new difficulty in the problem of finding the fluctuations of the heat, as this dependency affects the path integral formalism by introducing a new degree of freedom term in the stochastic action \cite{wio2013path, paraguassu2022heat2}.

However, in our analysis, we are interested in the difference of the kinetic energy within a given interval $t\in[0,\tau]$, which is given by
\begin{equation}
    \Delta K = m (v_{x,\tau}^2+v_{y,\tau}^2-v_{x,0}^2-v_{y,0}^2)/2.
\end{equation}
This is the quantity of interest in this article, and we will only refer to it as kinetic energy.

Despite the fact that the kinetic energy depends only on the initial and final conditions, we will see that this formula leads to non-trivial results. For instance, consider a free one-dimensional particle in an underdamped system. In this case, the heat is equal to the kinetic energy, and the distribution of kinetic energy follows a Bessel function of the second kind.  While for the 2D case, the heat follows a Laplace distribution, as demonstrated in \cite{paraguassu2023heat}.

\section{Distributions for the velocities}\label{probab}

Here, we will examine the distributions related to velocities, including the conditional probability distribution and its asymptotic behavior.

The conditional probability distribution offers information into the velocity of a particle, given its initial velocity $(v_{x,0}, v_{y,0})$. In this context, we derive the conditional probability distribution utilizing the path integral method. The specific calculations are outlined in detail in \ref{app}. For the final velocities $(v_{x,\tau}, v_{y,\tau})$, we obtain the following result
\begin{eqnarray}
    \fl P[v_{x,\tau},v_{y,\tau},\tau|v_{x,0},v_{y,0},0] \sim \exp\Bigg[{-\frac{m e^{-\frac{2 \gamma  \tau }{m}} \left(e^{\frac{2 \gamma  \tau }{m}}-1\right) \rm{csch}^2\left(\frac{\gamma\tau}{m}\right)}{8 T \left(\gamma ^2+\Gamma ^2\right)}}\times\nonumber\\
   \fl \Bigg(2 E_x (\gamma  v_{x,0}-\Gamma  v_{y,0})+2 E_y (\Gamma  v_{x,0}+\gamma  v_{y,0})+\left(\gamma ^2+\Gamma ^2\right) \left(v_{x,0}^2+v_{y,0}^2\right)\nonumber\\\fl
   +e^{\frac{2 \gamma  \tau }{m}} \left(2 E_x (\gamma  v_{x,\tau}-\Gamma  v_{y,\tau})+2 E_y (\Gamma  v_{x,\tau}+\gamma  v_{y,\tau})+\left(\gamma ^2+\Gamma ^2\right) \left(v_{x,\tau}^2+v_{y,\tau}^2\right)\right)\nonumber\\
   \fl +e^{\frac{\gamma\tau}{m}}\left(\cos \left(\frac{\tau}{\tau_\Gamma}\right) \left(-(v_{x,0}+v_{x,\tau}) (\gamma  E_x+\Gamma  E_y)+(v_{y,0}+v_{y,\tau}) (\Gamma  E_x-\gamma  E_y)\right.\right.\nonumber\\\fl\left.-\left(\gamma ^2+\Gamma ^2\right) (v_{x,0} v_{x,\tau}+v_{y,0} v_{y,\tau})\right)+(\gamma  E_x (v_{y,\tau}-v_{y,0})+\gamma  E_y (v_{x,0}-v_{x,\tau}) \nonumber\\\fl\left.\left.\left.+\Gamma E_x (v_{x,\tau}-v_{x,0})+\Gamma  E_y (v_{y,\tau}-v_{y,0})-\left(\gamma ^2+\Gamma ^2\right) (v_{x,\tau} v_{y,0}-v_{x,0} v_{y,\tau})\right)\sin \left(\frac{\tau}{\tau_\Gamma}\right)\right)\right],\label{fullprob}
\end{eqnarray}
where we omit the normalization constant, and write it in the \ref{app}. We define the frequency $\tau_\Gamma^{-1} = \Gamma/m$ which is the Larmor frequency \cite{liboff1966brownian,lemons1999brownian}. This is the conditional probability for the full dynamic (magnetic plus electric interaction). We can see that the magnetic field couples with the electric field, due products of $\Gamma E_i$ and the products with oscillations terms, like $ \cos{\left({ \tau}/{\tau_\Gamma}\right)}E_i,\ \sin{\left({ \tau}/{\tau_\Gamma}\right)}E_i$. These combinations will affect the kinetic energy of the particle. Furthermore, it is expected that the magnetic field alone would not be capable of affecting the particle's kinetic energy, as it does not perform work, a result we will observe next. Moreover, due to the coupling with the electric field, the magnetic field becomes influential in the transient and asymptotic regime.

The asymptotic time behavior of the distribution is given by
\begin{eqnarray}
\fl\lim_{\tau\rightarrow\infty}P[v_{x,\tau},v_{y,\tau},\tau|v_{x,0},v_{y,0}] = \frac{m}{2\pi T} \exp\left[-\frac{m }{2 T(\gamma^2+\Gamma^2)}\left(\left(v_{x,\tau}\gamma+E_x\right)^2+\left(v_{y,\tau}\gamma+E_y\right)^2\right)\right]\nonumber \\ \times \exp\left[-\frac{m \Gamma }{2 T(\gamma^2+\Gamma^2)}\left(v_{x,\tau}\left(v_{x,\tau}\Gamma+2E_y\right)+v_{y,\tau}\left(v_{y,\tau}\Gamma-2E_x\right)\right)\right].\label{equi3}
\end{eqnarray}
Now the velocity components become independent. This distribution is almost the free equilibrium distribution (if we look only to the quadratic velocity terms), but shifted by the electric field in combination with the strength of the magnetic field $\Gamma$. As a result, the particle's average velocity becomes proportional to the applied electric and magnetic fields. It is the equilibration with the static magnetic and electric field after long time. Unlike the free equilibrium distribution, which is just
\begin{equation}
    P_0(v_{x,0},v_{y,0})=\frac{m }{2 \pi  T}\exp\left({-\frac{m \left(v_{x,0}^2+v_{y,0}^2\right)}{2 T}}\right). \label{equi2}
\end{equation}
Nevertheless, this will be our initial state for the velocities, in all cases considered in the paper. Since we are going to see the subsequent effect of the Lorentz force in the free underdamped Brownian particle. 

Next, we will examine two specific cases: the dynamics purely driven by the electric field and the dynamics purely influenced by the magnetic field.

\subsection{Pure Electric Field}
Removing the magnetic field by $\Gamma \rightarrow 0$ in Eq.~\ref{fullprob} we decouple the velocities components, and find the conditional distribution 
\begin{eqnarray}
P[v_{x,\tau},v_{y,\tau},\tau|v_{x,0},v_{y,0},0]=P[v_{x,\tau},\tau|v_{x,0},0]P[v_{y,\tau},v_{y,0},0],
\end{eqnarray}
where
\begin{eqnarray}
  \fl  P[v_{y,\tau},\tau|v_{y,0},0] = \sqrt{\frac{m e^{\frac{2\gamma\tau}{m}} \left(\coth \left(\frac{\gamma\tau}{m}\right)-1\right)}{4\pi T}}\times\nonumber\\\exp \left(-\frac{m \left(\coth \left(\frac{\gamma\tau}{m}\right)-1\right) \left((E_y+\gamma  v_{y,0})-(E_y+\gamma  v_{y,\tau}) e^{\frac{\gamma\tau}{m}}\right)^2}{4 \gamma ^2 T}\right).\label{eleprob}
\end{eqnarray}
The distribution for the $x$-component is the same as the equation above, merely requiring the exchange of variables from $y$ to $x$.
Moreover, the asymptotic time limit is almost equal to the previous one, that is, the limit of $\tau \rightarrow \infty$ of Eq.~\ref{eleprob} is Eq.~\ref{equi3}, with $\Gamma \rightarrow0$.

\subsection{Pure Magnetic Field}
Making $E_x, E_y \rightarrow 0$ in Eq.~\ref{fullprob} we have the conditional distribution for the pure magnetic field case, that is
\begin{eqnarray}
    \fl P[v_{x,\tau},v_{y,\tau},\tau|v_{x,0},v_{y,0},0] = \frac{m}{2 \pi  T \left(1-e^{-\frac{2 \gamma  \tau }{m}}\right)} \times \label{pInt} \\\exp\left(- \left(\coth \left(\frac{\gamma\tau}{m}\right)-1\right)\frac{m  \left(v_{x,0}^2+v_{y,0}^2+e^{\frac{2\gamma\tau}{m}}\left(v_{x,\tau}^2+v_{y,\tau}^2\right)\right)}{4  T }\right)\times
\nonumber\\
    \fl \exp\left[-\frac{me^{\frac{\gamma\tau}{m}}\left(\coth \left(\frac{\gamma\tau}{m}\right)-1\right)}{2 T}\left( \sin \left(\frac{\tau }{\tau_\Gamma}\right) (v_{x,\tau} v_{y,0}-v_{x,0} v_{y,\tau})- \cos \left(\frac{\tau }{\tau_\Gamma}\right) (v_{x,0} v_{x,\tau}+v_{y,0} v_{y,\tau})\right)\right].\nonumber\label{magn}
\end{eqnarray}

Let's analyze the asymptotic behavior of the distribution given by Eq. \eref{pInt}. It can be observed that as $\tau$ approaches infinity, we have 
\begin{equation}
   \lim_{\tau\rightarrow\infty} P[v_{x,\tau}, v_{y,\tau}, \tau  | v_{x,0}, v_{y,0}] = \frac{m}{2\pi T}\exp\left(-m\frac{v_{x,\tau}^2+v_{y,\tau}^2}{2T}\right),\label{equi}
\end{equation} 
which is the Bolztmann equilibrium distribution. Besides the non-conservative force that comes from the magnetic field, the system reaches equilibrium for asymptotic times. This is already know in the literature, and is a version of the Bohr-Van Leeuwen Theorem for Brownian motion \cite{matevosyan_lasting_2023,pradhan_nonexistence_2010}.

We can analyze the distribution of the joint probability for $v_{x,\tau}$ and $v_{y,\tau}$. Lets assume that our system starts in equilibrium, meaning that the initial velocities obey the distribution in Eq.~\eref{equi2}. In doing so, we are interested in observing the effect of the static magnetic and electric fields on an initially free particle. By multiplying the initial velocities in Eq.~\eref{pInt} we have the joint distribution of Eq.~\eref{equi}. The static forces effect is to drive the particle towards a new static state, as described by the distribution in Eq.~\eref{equi3}. Consequently, we obtain a joint probability distribution that exhibits correlations between the initial and final velocities, along with temporal dependence. This results in the emergence of a transient regime in the particle's dynamics.

What's intriguing is that even if the initial velocities follow Eq.~\eref{equi3} (meaning the particle starts in the asymptotic state due to the Lorentz force), we still observe a transient regime. This transient behavior arises primarily from the conditional distributions, which introduce temporal dependencies into the joint distribution, causing the distribution of kinetic energy to evolve dynamically rather than remaining static.

\section{Characteristic Function and Central Moments}\label{Cfun}
The characteristic function is the Fourier transform of the probability distribution and can be calculated using the formula
\begin{equation}
    Z(\lambda) = \int dv_{x,0}dv_{y,0} dv_{x,\tau}dv_{y,\tau} P\left(v_{x,\tau},v_{y,\tau},v_{x,0},v_{y,0}\right) \exp\left(-i\lambda \Delta K\right)\label{zldef}
\end{equation}
where the joint distribution is constructed from the multiplication of Eq.~\eref{fullprob} with $P_0(v_{x,0},v_{y,0})$ (see Eq.~\eref{equi2}).

We will first analyze the pure magnetic case, where we find that the magnetic field alone cannot affect the fluctuations of the kinetic energy.

\subsection{Magnetic Field does not affect the kinetic distribution}
By removing the electric forces, we find the characteristic function of the pure magnetic field case, and find that it is the free particle case, that is, substituting Eq.~\eref{pInt} in Eq.~\eref{zldef}, we have
 \begin{equation}
     Z_B(\lambda) = \frac{1}{\lambda ^2 T^2 \left(1-e^{-\frac{2 \gamma  \tau }{m}}\right)+1}.
 \end{equation}
where $B$ is to denote that we have only the magnetic force acting. It is important to notice that the absence of $\Gamma$ comes from the Gaussian integration of Eq.~\eref{zldef}. By doing the Gaussian integration the final velocities or the initial ones, the result becomes independent of $\Gamma$ due the way that the velocities combine in Eq.~\eref{magn}.  With this characteristic function, the mean and skewness are null, meaning that the kinetic distribution is centered at the origin and symmetric in its argument. Moreover, the variance is given by 
 \begin{equation}
     \sigma^2_{\Delta K} = 2 T^2 \left(1-e^{-\frac{2 \gamma  \tau }{m}}\right),
 \end{equation}
 and is positive and monotonic increasing in time. For the excess of kurtosis we find the simple result of $\kappa = 3$. Showing that the distribution is leptokurtic, as already know in \cite{paraguassu2022heat2}. Since the kinetic energy is independent of the magnetic field, this result is the same for the heat in the underdamped free particle case \cite{paraguassu2022heat2}.

\subsection{Full case}
After integrating Eq.~\eref{zldef} with the Lorentz force (electric + magnetic field), we obtain
\begin{eqnarray}
\fl Z(\lambda)=  \frac{\left(\coth \left(\frac{\gamma\tau}{m}\right)+1\right) }{\coth \left(\frac{\gamma\tau}{m}\right)+2 \lambda ^2 T^2+1}\times\nonumber\\\exp \left(\frac{\lambda  m \left(E_x^2+E_y^2\right) (\lambda  T+i) \rm{csch}\left(\frac{\gamma\tau}{m}\right) \left(\cos \left(\frac{\tau}{\tau_\Gamma}\right)-\cosh \left(\frac{\gamma\tau}{m}\right)\right)}{\left(\gamma ^2+\Gamma ^2\right) \left(\coth \left(\frac{\gamma\tau}{m}\right)+2 \lambda ^2 T^2+1\right)}\right).\label{fullcharac}
\end{eqnarray}
One important result is in the asymptotic time. Where the characteristic function becomes
\begin{equation}
     Z(\lambda)= \frac{1}{\lambda ^2 T^2+1}\exp \left(-\frac{\lambda  m \left(E_x^2+E_y^2\right)}{2 \left(\gamma ^2+\Gamma ^2\right) (\lambda  T-i)}\right),
\end{equation}
which depends on the magnetic field. Therefore, due to the electric field, not only does the magnetic field become relevant, but it also has a lasting effect on the kinetic energy distribution.

\begin{figure}
    \centering
    \includegraphics[width=15.5cm]{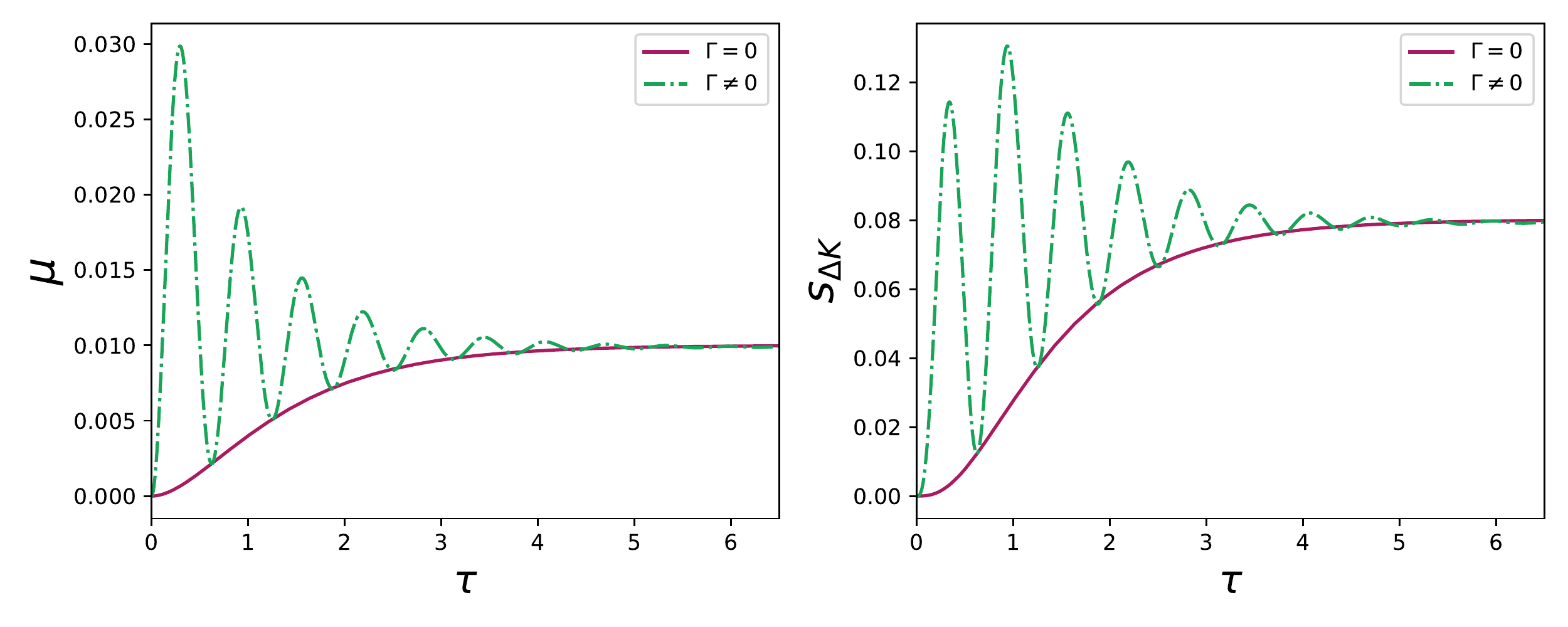}
    \caption{Left) Mean over time for: the pure electric case with $E_x=E_y=0.1$ and $\Gamma = 0$ (solid line), the full case with $\Gamma=10$ and $E_x=E_y=1$ (dot dashed line). Right) Skewness over time for: the pure electric case with $E_x=E_y=0.1$ and $\Gamma = 0$ (solid line), the full case with $\Gamma=10$ and $E_x=E_y=1$ (dashed line). All the remaining constants are set to one.  }
    \label{avgskew}
\end{figure}

By deriving the characteristic function, Eq.~\eref{fullcharac}, we can obtain the moments and thereof the central moments \cite{paraguassu2022heat2}. The first one is the mean, which is
\begin{equation}
   \mu= \frac{m \left(E_x^2+E_y^2\right) \left(\coth \left(\frac{\gamma\tau}{m}\right)-\rm{csch}\left(\frac{\gamma\tau}{m}\right) \cos \left(\frac{\tau}{\tau_\Gamma}\right)\right)}{\left(\gamma ^2+\Gamma ^2\right) \left(\coth \left(\frac{\gamma\tau}{m}\right)+1\right)}.
\end{equation}
We can observe from the mean that the average kinetic energy depends quadratically on the electric fields, implying that the sign of these quantities does not matter. Additionally, the dependence on the magnetic field is merely an oscillation and only exists due to the presence of the electric field. We plot the mean in the left of \fref{avgskew}. Where we can see that the mean is positive and increase in time, saturating at a fixed value of
\begin{equation}
    \lim_{\tau\rightarrow\infty} \mu = \frac{m \left(E_x^2+E_y^2\right)}{2 \left(\gamma ^2+\Gamma ^2\right)}.
\end{equation}
Note that the asymptotic limit carries a dependence on the magnetic interaction by $\Gamma$. Showing that the electric field allows the effect of the magnetic field to persist at long times.

The next central moment is the variance, given by
\begin{equation}
  \fl  \sigma^2_{\Delta K} = \frac{2 T e^{-\frac{\gamma\tau}{m}} \left(2 T \left(\gamma ^2+\Gamma ^2\right) \sinh \left(\frac{\gamma\tau}{m}\right)-m \left(E_x^2+E_y^2\right) \left(\cos \left(\frac{\tau}{\tau_\Gamma}\right)-\cosh \left(\frac{\gamma\tau}{m}\right)\right)\right)}{\gamma ^2+\Gamma ^2},
\end{equation}
which is always positive and increasing in time. We plotted in the left of \fref{varkurt}. Its asymptotic time limit is given by
\begin{equation}
  \lim_{\tau\rightarrow\infty} \sigma^2_{\Delta K} = \left(T\frac{m \left(E_x^2+E_y^2\right)}{\gamma ^2+\Gamma ^2}+2 T^2\right),
\end{equation}
which also depends on the magnetic field constant $\Gamma$.

The skewness informs about the asymmetry of the distribution, is defined as the third central moment, and for our system, due the electric field, is non-null, and given by
\begin{eqnarray}
    \fl s_{\Delta K} =\frac{m \left(E_x^2+E_y^2\right) \left(\coth \left(\frac{\gamma\tau}{m}\right)-\rm{csch}\left(\frac{\gamma\tau}{m}\right) \cos \left(\frac{\tau}{\tau_\Gamma}\right)\right)}{\left(\gamma ^2+\Gamma ^2\right)^3 \left(\coth \left(\frac{\gamma\tau}{m}\right)+1\right)^3}\times\nonumber \\\fl \Bigg[m^2 \left(E_x^2+E_y^2\right)^2 \left(\coth \left(\frac{\gamma\tau}{m}\right)-\rm{csch}\left(\frac{\gamma\tau}{m}\right) \cos \left(\frac{\tau}{\tau_\Gamma}\right)\right)^2+2T\left(\gamma ^2+\Gamma ^2\right)\left(8 T \left(\gamma ^2+\Gamma ^2\right)\right.\nonumber\\\left.\fl+m \left(E_x^2+E_y^2\right) \left(\coth \left(\frac{\gamma\tau}{m}\right)-\rm{csch}\left(\frac{\gamma\tau}{m}\right) \cos \left(\frac{\tau}{\tau_\Gamma}\right)\right)\right)\left(\coth \left(\frac{\gamma\tau}{m}\right)+1\right)\Bigg].
\end{eqnarray}
Just like the mean, the skewness also increases with time and exhibits oscillations due to the magnetic field. It is positive, indicating that positive values are more probable. By looking at the right of \fref{avgskew}, we can see that it has a peak due to the oscillations, which means that having the electric field together with the magnetic field has a non-monotonic behavior \cite{manikandan2022nonmonotonic}. In contrast to the case where only the magnetic field is present, and the skewness is zero, when we use the electric field, we can increase the particle's velocity due to the magnetic field. While for the pure electric case, the evolution of the moments is monotonic, as we will see in the next subsection.

To analyze the extreme values of the distribution, we calculate the excess of kurtosis, which is the kurtosis (fourth central moment) minus 3 \cite{westfall2014kurtosis}. The excess of kurtosis is
\begin{eqnarray}
   \fl \kappa_{\Delta K} = \frac{12 T \left(\gamma ^2+\Gamma ^2\right) \left(m \left(E_x^2+E_y^2\right) \left(\coth \left(\frac{\gamma\tau}{m}\right)-\rm{csch}\left(\frac{\gamma\tau}{m}\right) \cos \left(\frac{\tau}{\tau_\Gamma}\right)\right)+T \left(\gamma ^2+\Gamma ^2\right)\right)}{\left(m \left(E_x^2+E_y^2\right) \left(\coth \left(\frac{\gamma\tau}{m}\right)-\rm{csch}\left(\frac{\gamma\tau}{m}\right) \cos \left(\frac{\tau}{\tau_\Gamma}\right)\right)+2 T \left(\gamma ^2+\Gamma ^2\right)\right)^2}.\nonumber \\
\end{eqnarray}
The found excess of kurtosis is always greater than zero, meaning that the distribution is leptokurtic \cite{westfall2014kurtosis}, which means that the probability distribution has heavier tails compared to the normal distribution. In other words, it indicates that the distribution has more extreme values and a higher probability of observing outliers or extreme events. Note that the kurtosis decreases because there are fewer chances of rare events occurring as time passes.
Moreover, it is interesting to note that leptokurtic behavior is observed in the statistics of heat in various systems \cite{paraguassu2022effects,paraguassu2023heat,paraguassu2023heat2,paraguassu2022heat2}.

\subsection{Pure Electric Field}
\begin{figure}
    \centering
    \includegraphics[width=15.5cm]{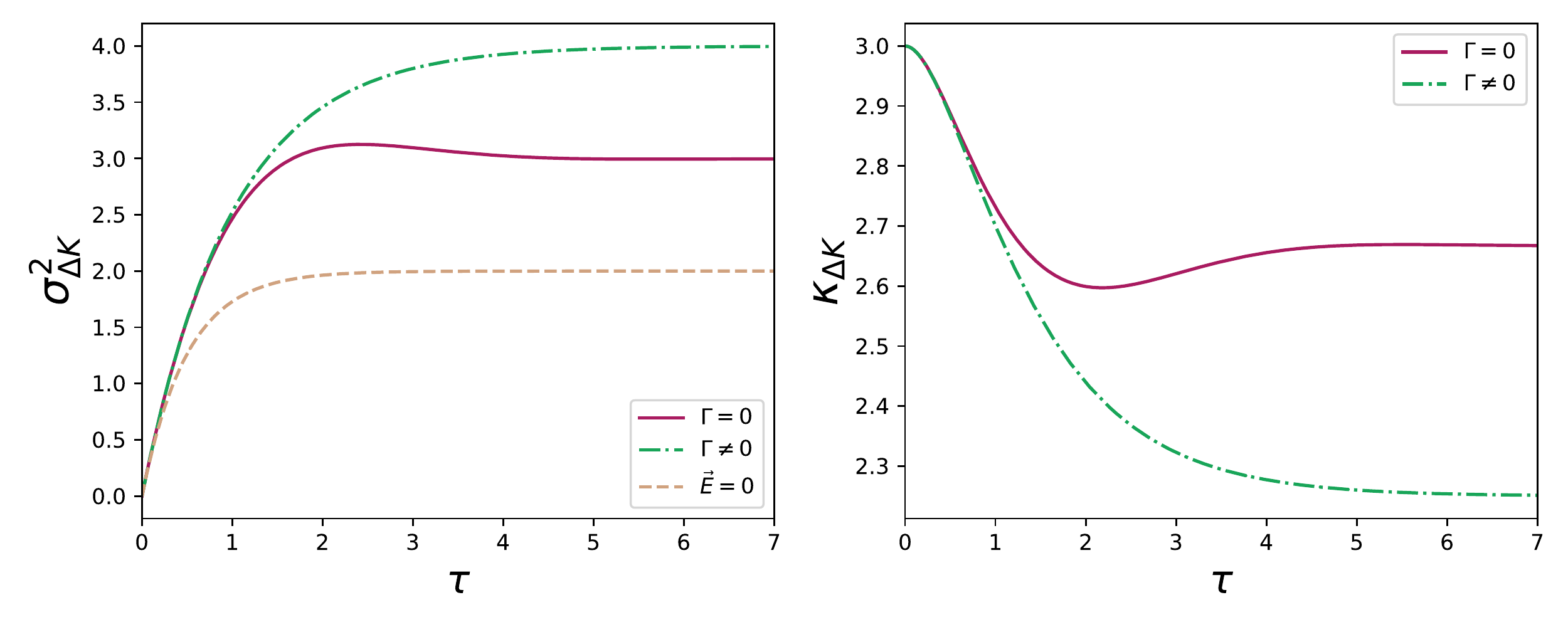}
    \caption{Left) Variance over time for: the full case (dot dashed line) with $\Gamma=10$ and $E_x=E_y=1$, the pure electric case with $E_x=E_y=0.1$ and $\Gamma = 0$ (solid line), and the pure magnetic with $E_x=E_y=0.1$ and $\Gamma=1$. Right) Excess of kurtosis over time for: the full case (dot dashed line) with $\Gamma=10$ and $E_x=E_y=1$, and the pure electric case with $E_x=E_y=0.1$ and $\Gamma = 0$ (solid line). The pure magnetic case is constant with $\kappa=3$. All the remaining constants are set to one.}
    \label{varkurt}
\end{figure}

Ignoring the magnetic field by taking $\Gamma \rightarrow 0$. We remove the oscillations in the moments, since it is the terms $\sim \cos\left(\tau/\tau_\Gamma\right)$ that induces the oscillations. The characteristic function  becomes
\begin{equation}
   \fl Z_E(\lambda)= \frac{\left(\coth \left(\frac{\gamma\tau}{m}\right)+1\right) }{\coth \left(\frac{\gamma\tau}{m}\right)+2 \lambda ^2 T^2+1}\exp \left(-\frac{\lambda  m \left(E_x^2+E_y^2\right) (\lambda  T+i) \left(e^{\frac{\gamma\tau}{m}}-1\right)^2}{2 \gamma ^2 \left(\left(\lambda ^2 T^2+1\right) e^{\frac{2 \gamma  \tau }{m}}-\lambda ^2 T^2\right)}\right),
\end{equation}
where the subscript $E$ is to denote that we are working with the pure electric case. Its asymptotic value is given by
\begin{equation}
   \fl \lim_{\tau\rightarrow\infty}Z_E(\lambda)=  \frac{1}{\lambda ^2 T^2+1}\exp\left({-\frac{\lambda  m \left(E_x^2+E_y^2\right)}{2 \gamma ^2 (\lambda  T-i)}}\right).
\end{equation}
The behavior of the distribution will be almost the same as that of the previous central moments, as we only need to let $\Gamma \rightarrow 0$ in the previous central moments. Consequently, we choose to analyze these moments only graphically, represented by the solid lines in  \fref{avgskew} and \fref{varkurt}.

We can observe, then, that unlike the case with the magnetic field, the purely electric case does not exhibit oscillations in quantities such as mean, variance, skewness, and kurtosis. Note that we have chosen different values for comparison in the figures \ref{avgskew}, \ref{varkurt} and used a weak electric field to be in the same scale of the moments in the case with the magnetic and electric field. Therefore, the electric field alone, produces higher values for the kinetic energy central moments than the combined case, increasing the fluctuations of positive values in the kinetic energy increment. Nevertheless, by combining the electric field with the magnetic field, we enable the system to experience a permanent effect of the magnetic field. In contrast to the pure magnetic case, which does not affect the fluctuations of kinetic energy due to the Bohr-Van Leuween theorem.

Next, we will examine the probability distributions for the increment of kinetic energy in the cases studied here.

\section{Kinetic energy distributions}\label{Dist}
Now let's calculate the distribution of the kinetic energy by taking the Fourier transform of the characteristic function.

\begin{figure}
    \centering
    \includegraphics[width=15cm]{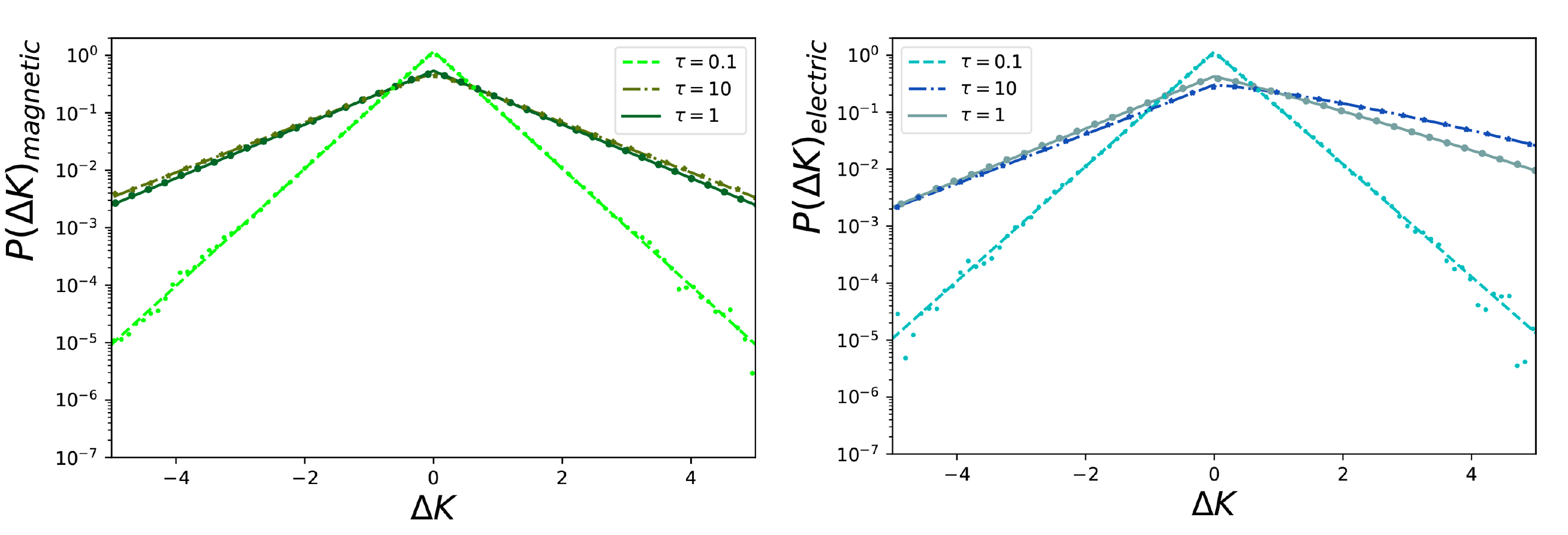}
    \caption{Left) Kinetic energy distribution in the pure magnetic case  for $\tau=0.1,1,10$. Note that, despite the distribution at $\tau=1$ and $\tau=10$ be very similar, the latter has more fluctuations for values far from the origin. Right) Kinetic energy distribution in the pure electric case  for $\tau=0.1,1,10$. Note that, the distribution is asymmetric, with more area in the positive side of $\Delta K$. We plot these results with values of $\gamma=T=m=1$. The dots are the numerical simulations for different times}
    \label{plot}
\end{figure}

We start with the pure magnetic case, since the distribution can be achieved analytically. The result is the distribution of the kinetic energy in the free case, that is
\begin{equation}
   P_B(\Delta K) =  \frac{1 }{2T \sqrt{1-e^{-\frac{2 \gamma  \tau }{m}}}}\exp\left({-\frac{| \Delta K| }{T \sqrt{1-e^{-\frac{2 \gamma  \tau }{m}}}}}\right),\label{kinedist}
\end{equation}
and is a transient distribution with asymptotic time limit being a simple exponential distribution.
\begin{equation}
    \lim_{\tau \rightarrow \infty} P_B(\Delta K) = \frac{e^{-\frac{|\Delta K|}{T}}}{2T}.
\end{equation}
We plot this distribution and compare it with numerical simulations considering the dynamics in Eq~\eref{langevin}. The resulting plot is shown in the left of figure \ref{plot}. It is evident that the distribution exhibits increasing fluctuations as time progresses. This distribution is symmetric, and centered in the origin, as expected since the magnetic field alone is not capable of changing the kinetic energy.

For the underdamped free particle, the calculation of the kinetic energy distribution has already been done in \cite{paraguassu2022effects} (see Eq.(27) of \cite{paraguassu2022effects}) and it was shown to be equal to the heat, as the particle is free. Therefore, our result for the kinetic energy of an underdamped particle with a magnetic force acting on it yields exactly the same distribution. This implies that the presence of a static magnetic field, which affects the dynamics of the particle, does not have any impact on its kinetic energy behavior. Despite the fact that the joint distribution depends on $\Gamma$, after integrating, the left results are independent. However, in the free case, $v_x$ and $v_y$ are uncorrelated, and here for the magnetic field case they were correlated, nevertheless, the result still be the same.

The independence of the magnetic field in the kinetic energy distribution can be viewed as a consequence of the Bohr-Van Leeuwen theorem, which has been previously explored in the context of stochastic systems \cite{matevosyan_lasting_2023,pradhan_nonexistence_2010}. As the magnetic field does not perform any work, the distribution of kinetic energy remains unaffected.

For the other two cases, the pure electric and the full case, we are not able to obtain analytical expressions for the distribution. However we can calculate the integrals numerically and compare with numerical simulations of the Langevin equation. The pure electric case is plotted in the right of \fref{plot}. Right away, we can observe that the distribution is asymmetric, with a higher probability of positive values occurring. Clearly, the area for $\Delta K>0$ is larger than for $\Delta K<0$. This aligns with the expectations from the skewness analysis. Additionally, the fluctuations in positive values increase as time passes.

\begin{figure}
    \centering
    \includegraphics[width=12cm]{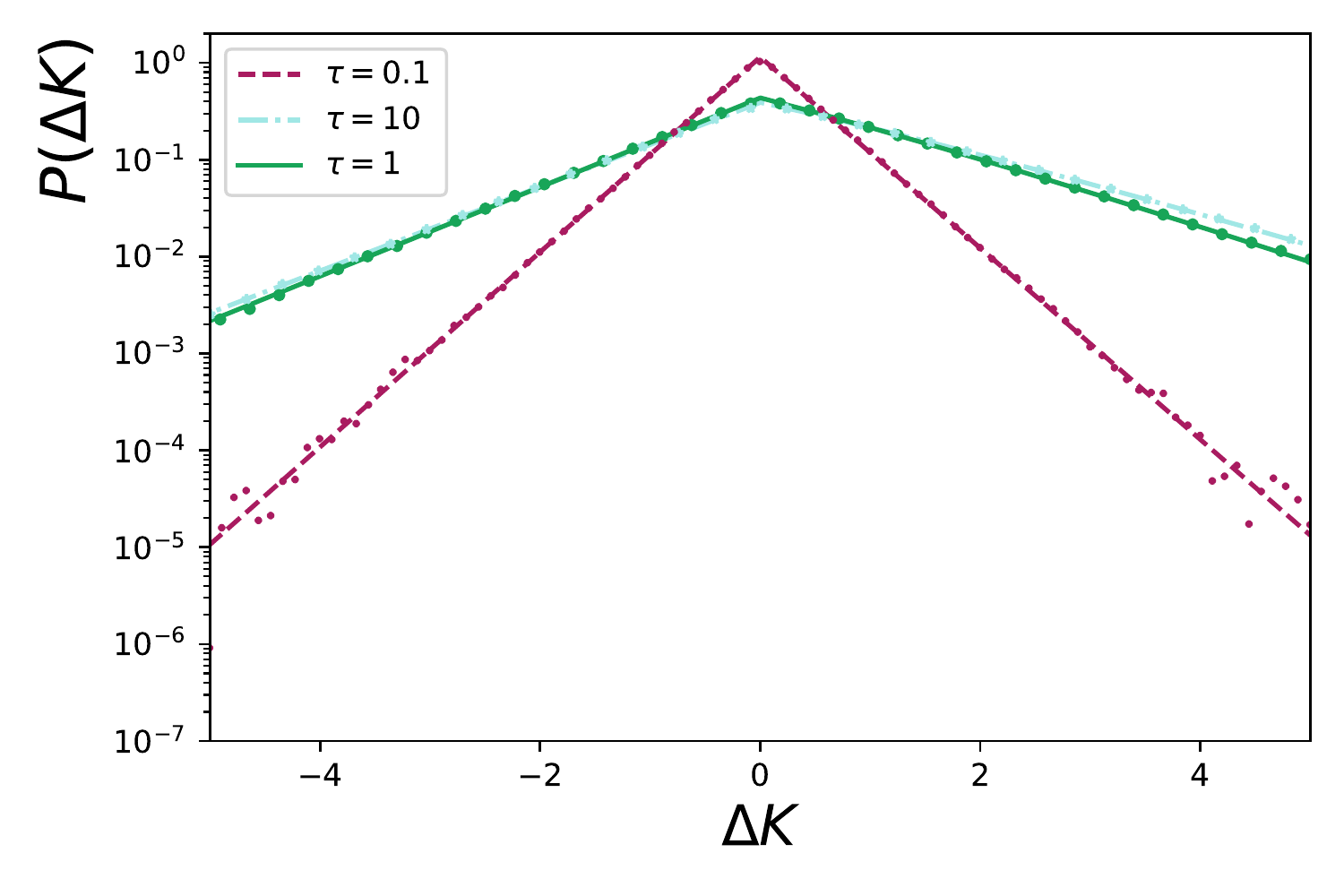}
    \caption{Distribution with both magnetic and electric field for different times. All remaining constants are set to one. The asymmetry is evident in the right side, were positive values are more probable.}
    \label{fplot2}
\end{figure}

For the case with both fields, we plot the results in \fref{fplot2}, which show similarities with the purely electric case, as the effect of the magnetic field is modest. Although there remains some asymmetry, it is less pronounced and less noticeable. This finding is consistent with the skewness measurement, as shown in \fref{avgskew}, where we needed to use a very weak electric field to make a comparison with the values obtained in the presence of the magnetic field.

\section{Conclusion}\label{conclusion}
Despite being a seemingly straightforward quantity, kinetic energy exhibits intriguing statistical behavior in the realm of stochastic systems. In this study, we investigate a two-dimensional Brownian particle initially at free equilibrium, subjected to a perpendicular static magnetic field and electric fields in both directions of motion, which correspond to the Lorentz force. The presence of the magnetic field affects the dynamics of the particle's velocity, leading to the emergence of correlated components in the conditional distributions. In contrast, the electric field continually propels the particle.

By employing path integrals, we derive the conditional probability distribution for three cases: the combination of magnetic and electric field, the pure magnetic and the pure electric field. We found different dynamics in each case. For the pure magnetic case we observe that the system attains free equilibrium, not depending on the magnetic field, despite the non-conservative force stemming from the magnetic field. For the full case, by using an electric field together with the magnetic field, the system attains a steady distribution, depending on the parameters of both fields.

With the dynamics characterized, we proceed to examine the kinetic energy increment in a given interval. We calculate its characteristic function, central moments, and the corresponding distribution for the three cases considered.
We find that the kinetic energy for the pure magnetic case follows an exponential distribution, independent of the magnetic field parameter. A consequence of the Bohr-Van Leeuwen theorem. This intriguing finding leads us to the conclusion that, despite the dynamics being affected, the kinetic energy remains unaffected.

Notably, our results with the pure magnetic case align with the results of an underdamped free particle case in two dimensions. The kinetic energy distribution for the two systems with different dynamics are the same. While by using an electric field, the kinetic energy is affected by the magnetic field, since the coupling between the two field creates a permanent effect of the magnetic field, responsible by oscillations of the statistical behavior of the kinetic energy. That is, a non-monotonic behavior in the central moments of the kinetic energy. Moreover, due the electric field, the kinetic energy acquires a bias, with a tendency for more probable positive values. This insight emphasizes the crucial role played by the electric field in shaping the kinetic energy distribution.

In summary, our study focus on the intriguing statistical behaviors of kinetic energy in the presence of magnetic and electric fields, uncovering various dynamic regimes and offering valuable insights into the interplay between these fields. These findings could have potential applications in the diffusion of electrons and ions in a plasma where the studied model is used. 

\appendix
 \section{Path Integral Calculation}\label{app}

 The conditional probability distribution is given by the formula
 \begin{equation}
     P[v_{x,\tau},v_{y,\tau},\tau|v_{x,0},v_{y,0},0] = \int \mathcal{D}v_x\int\mathcal{D}v_y \exp\left(-\frac{1}{4\gamma T}S[v_x,v_y]\right)
 \end{equation}
This expression arises from the statistical properties of the white noise $\eta(t)$ \cite{wio2013path}. And in our case, similar path integral was already calculated in \cite{kokiantonis1985propagator,cheng1984propagator,chatterjee2011single}. The action is quadratic in its arguments, and given by
\begin{equation}
   \fl S[v_x,v_y]= \int_0^\tau\left(m\dot v_x+\gamma v_x +\Gamma v_y+E_x\right)^2dt+\int_0^\tau\left(m\dot v_y+\gamma v_y -\Gamma v_x+E_y\right)^2dt,\label{lagran}
\end{equation}
implying that the result can be computed by considering only the deterministic path, which corresponds to evaluating the action at its extrema \cite{wio2013path, paraguassu2022effects, chaichian2018path}. Thus, we have:
\begin{equation}
P[v_{x,\tau},v_{y,\tau},\tau|v_{x,0},v_{y,0},0] \sim \exp\left(-\frac{1}{4\gamma T}S_{d}[v_x,v_y]\right),\label{prePin}
\end{equation}
The normalization factor can be calculated subsequently. The $S_d$ action represents the action evaluated at the solution of the Euler-Lagrange equations for the velocities, given by:
\begin{eqnarray}
 \gamma  E_x+\Gamma  E_y+2 \Gamma  m \dot v_y(t)+\left(\gamma ^2+\Gamma ^2\right) v_x(t)&=&m^2 \ddot v_x(t), \\  \gamma  E_y+\Gamma  E_x-2 \Gamma  m \dot v_x(t)+\left(\gamma ^2+\Gamma ^2\right) v_y(t)&=&m^2 \ddot v_y(t).
\end{eqnarray}
Solving both equation with the boundary conditions $v_i(0)=v_{i,0}, v_i(\tau)=v_{i,\tau} $ we find the deterministic paths,
\begin{eqnarray}
    \fl v_{x}(t) = \frac{\left(\coth \left(\frac{\gamma  \tau }{m}\right)-1\right) e^{-\frac{\gamma  (3 t+\tau )}{m}}}{2 \left(\gamma ^2+\Gamma ^2\right)}\Bigg((\gamma  E_x+\Gamma  E_y) \left(-\left(e^{\frac{3 \gamma  (t+\tau )}{m}}-e^{\frac{\gamma  (3 t+\tau )}{m}}\right)\right)+ \nonumber \\\fl -\left(e^{\frac{2 \gamma  (t+\tau )}{m}}-e^{\frac{2 \gamma  (2 t+\tau )}{m}}\right) \cos \left(\frac{\Gamma  (t-\tau )}{m}\right) \left(\gamma  E_x+\Gamma  E_y+v_{x,\tau} \left(\gamma ^2+\Gamma ^2\right)\right)+ \nonumber \\ \fl+\left(e^{\frac{\gamma  (4 t+\tau )}{m}}-e^{\frac{\gamma  (2 t+3 \tau )}{m}}\right)\left(\cos \left(\frac{\Gamma  t}{m}\right) \left(-\gamma  E_x-\Gamma  E_y-v_{x,0} \left(\gamma ^2+\Gamma ^2\right)\right)\right. \nonumber \\ \fl\left.+\sin \left(\frac{\Gamma  t}{m}\right) \left(-\left(-\Gamma  E_x+\gamma  E_y+v_{y,0} \left(\gamma ^2+\Gamma ^2\right)\right)\right)\right) \nonumber \\ \fl-\left(e^{\frac{2 \gamma  (t+\tau )}{m}}-e^{\frac{2 \gamma  (2 t+\tau )}{m}}\right) \sin \left(\frac{\Gamma  (t-\tau )}{m}\right) \left(-\Gamma  E_x+\gamma  E_y+v_{y,\tau} \left(\gamma ^2+\Gamma ^2\right)\right)\Bigg),
\end{eqnarray}
\begin{eqnarray}
   \fl v_y(t)= \frac{\left(\coth \left(\frac{\gamma  \tau }{m}\right)-1\right) e^{-\frac{\gamma  (3 t+\tau )}{m}}}{2 \left(\gamma ^2+\Gamma ^2\right)}\Bigg(\left(e^{\frac{\gamma  (3 t+\tau )}{m}}-e^{\frac{3 \gamma  (t+\tau )}{m}}\right) (\gamma  E_y-\Gamma  E_x)+ \nonumber \\ \fl -\left(e^{\frac{2 \gamma  (t+\tau )}{m}}-e^{\frac{2 \gamma  (2 t+\tau )}{m}}\right) \cos \left(\frac{\Gamma  (t-\tau )}{m}\right) \left(\left(\gamma ^2+\Gamma ^2\right) v_{y, \tau }-\Gamma  E_x+\gamma  E_y\right)+ \nonumber \\\fl +\left(e^{\frac{\gamma  (4 t+\tau )}{m}}-e^{\frac{\gamma  (2 t+3 \tau )}{m}}\right)\left(\cos \left(\frac{\Gamma  t}{m}\right) \left(-\left(\gamma ^2+\Gamma ^2\right) v_{y, 0}+\Gamma  E_x-\gamma  E_y\right)\right.+ \nonumber \\ \fl\left.+\sin \left(\frac{\Gamma  t}{m}\right) \left(\left(\gamma ^2+\Gamma ^2\right) v_{x, 0}+\gamma  E_x+\Gamma  E_y\right)\right)\bigg) + \nonumber \\ \fl+\left(e^{\frac{2 \gamma  (t+\tau )}{m}}-e^{\frac{2 \gamma  (2 t+\tau )}{m}}\right) \sin \left(\frac{\Gamma  (t-\tau )}{m}\right) \left(\left(\gamma ^2+\Gamma ^2\right) v_{x\, \tau }+\gamma  E_x+\Gamma  E_y\right)\Bigg).
\end{eqnarray}
Now we just need to substitute these equations in Eq.~\eref{lagran} and calculate the time integral. After doing this, we obtain the action in its extrema, Eq.~\eref{prePin}. The result is the argument of the exponential in Eq.~\eref{fullprob}. What is left is the normalization constant, which can be calculated by implying $\int P[\dots] dv_{y,\tau}dv_{x,\tau}=1$. After doing this, the normalization is
\begin{eqnarray}
  \mathcal{N}= \frac{m}{4\pi T} \left(1+\coth\left(\frac{\gamma\tau}{m}\right)\right).
\end{eqnarray}
Therefore we have the complete distribution. 

 \section*{Acknowledges}
 The author would like to thank C. M. Daher, T. Guerreiro, W. Morgado for useful discussions. 
This work is supported by the Brazilian agencies CNPq, CAPES, and FAPERJ. P.V.P. would like to thank FAPERJ for his previous PhD fellowship. This study was financed in
part by Coordenação de Aperfeiçoamento de Pessoal de Nível
Superior - Brasil (CAPES) - Finance Code 001.
\section*{References}

\bibliographystyle{name}
\bibliography{name}

\end{document}